\documentclass{IEEEtran}
\usepackage{cite}
\usepackage{amsmath,amssymb,amsfonts}
\usepackage{algorithmic}
\usepackage{graphicx}
\usepackage{textcomp}

\usepackage[
  colorlinks=true,
  linkcolor=blue,
  citecolor=blue,
  urlcolor=blue
]{hyperref}

\ifCLASSOPTIONcompsoc
 \usepackage[caption=false,font=normalsize,labelfont=sf,textfont=sf, subrefformat=parens, labelformat=parens]{subfig}
\else
 \usepackage[caption=false,font=footnotesize, subrefformat=parens, labelformat=parens]{subfig}
\fi

\usepackage{amssymb}   
\usepackage{svg}
\usepackage{siunitx}
\usepackage[capitalise]{cleveref}
\crefformat{equation}{(#2#1#3)}
\crefmultiformat{equation}{(#2#1#3),}%
{ (#2#1#3)}{, (#2#1#3)}{ ,(#2#1#3)}
\crefrangeformat{equation}{(#3#1#4) - (#5#2#6)}
\Crefformat{equation}{Equation~(#2#1#3)}
\Crefmultiformat{equation}{Equations~(#2#1#3),}%
{ (#2#1#3)}{, (#2#1#3)}{ ,(#2#1#3)}
\Crefrangeformat{equation}{Equations~(#3#1#4) - (#5#2#6)}
\Crefmultiformat{figure}{Fig.~(#2#1#3),}%
{ (#2#1#3)}{, (#2#1#3)}{ ,(#2#1#3)}
\crefname{subsection}{subsection}{subsections}
\AtBeginDocument{\crefformat{subsection}{#2Subsection~#1#3}}

\usepackage{comment}

\usepackage{booktabs}
\usepackage{tabulary}

\allowdisplaybreaks  
\def\BibTeX{{\rm B\kern-.05em{\sc i\kern-.025em b}\kern-.08em
    T\kern-.1667em\lower.7ex\hbox{E}\kern-.125emX}}
\begin{document}
\title{Expressivity of Programmable-Metasurface-Based Physical Neural Networks: Encoding Non-Linearity, Structural Non-Linearity, and Depth}
\author{Cheima Hammami, Luc Le Magoarou,~\IEEEmembership{Member, IEEE}, Christos Monochristou, David González-Ovejero,~\IEEEmembership{Senior Member, IEEE}, Ali Momeni, Romain Fleury, and Philipp del Hougne, \IEEEmembership{Member, IEEE}
\thanks{
C.~Hammami, C.~Monochristou, D.~González-Ovejero and P.~del~Hougne are with Univ Rennes, CNRS, IETR - UMR 6164, F-35000, Rennes, France. P.~del~Hougne is also with the Department of Electronics and Nanoengineering, Aalto University, 00076 Espoo, Finland (e-mail: cheima.hammami@univ-rennes.fr; christos.monochristou@univ-rennes.fr; david.gonzalez-ovejero@univ-rennes.fr; philipp.del-hougne@univ-rennes.fr).
}
\thanks{L.~Le Magoarou is with INSA Rennes, CNRS, IETR - UMR 6164, F-35000, Rennes, France (e-mail: luc.le-magoarou@insa-rennes.fr).
}
\thanks{A.~Momeni and R.~Fleury are with the Laboratory of Wave Engineering, School of Engineering, École Polytechnique Fédérale de Lausanne, 1015 Lausanne, Switzerland (e-mail: ali.momeni@epfl.ch; romain.fleury@epfl.ch).
}
\thanks{\textit{(Corresponding Author: Philipp del Hougne.)}}
\thanks{This work was supported in part by the Nokia Foundation (project 20260028), the ANR France 2030 program (project ANR-22-PEFT-0005), and the ANR PRCI program (project ANR-22-CE93-0010).}
}

\maketitle

\begin{abstract}
Wave-based signal processing conventionally encodes input data into the input wavefront, making it challenging to implement non-linear operations.
Programmable wave systems enable an alternative approach: encoding the input data into the scattering properties of tunable components. With such structural input encoding, two potentially non-linear mappings are involved: first, from the input data to the tunable components' scattering characteristics, and, second, from these scattering characteristics to the output wavefront. In this paper, we systematically examine the expressivity of a wave-based physical neural network (WPNN) with structural input encoding. Our analysis is based on a physics-consistent multiport-network model of a compact D-band rich-scattering cavity parametrized by a 100-element programmable metasurface. We separately control encoding non-linearity, structural non-linearity, and network depth in order to examine their interplay, considering a controlled scalar regression task.
With phase encoding and strong inter-element mutual coupling (MC), both aforementioned mappings are strongly non-linear and the WPNN performs very well even with a single layer. We further observe that additional layers can partially compensate for weak inter-element MC. In addition, we demonstrate that WPNN depth can improve expressivity even when it is not associated with an increase in trainable weights. Altogether, our results provide a physics-consistent picture of how encoding choice, MC strength, and depth jointly govern the expressive power of PM-based WPNNs, informing design choices for future experimental implementations of WPNNs.
\end{abstract}


\section{Introduction}
\label{sec:introduction}

Programmable wave-domain computing (pWDC), i.e., off-loading specific-purpose computations from conventional electronic processors to the wave domain where signals are directly processed through reconfigurable wave–matter interactions, is emerging as a technological enabler to address challenges of all-digital signal processing related to metrics such as cost, energy consumption, latency, or footprint~\cite{pWDCperspective}. 
Signal processing in the wave domain has a long tradition, but could not compete with rapidly improving general-purpose electronic processors for a long time. Recent advances in the areas of metamaterial engineering and wave control, combined with the emergence of compute-hungry artificial intelligence, have sparked renewed interest in specific-purpose wave-domain computing across the electromagnetic spectrum~\cite{zangeneh2021analogue,mcmahon2023physics,abou2025programmable,pWDCperspective}. The hope of pWDC is to combine intrinsic benefits of wave-domain processing (e.g., related to bandwidth, parallelism, and power consumption~\cite{mcmahon2023physics}) with the wave-domain flexibility offered by programmable metasurfaces (PMs) to deterministically adjust the implemented operation during runtime~\cite{pWDCperspective}.

A fundamental challenge in pWDC lies in understanding the expressivity of the available reconfigurable wave-domain processor, i.e., how rich a set of functions the system can represent. Recent work evaluates upper bounds on the fidelity with which an experimentally given pWDC system can synthesize a desired \textit{linear} operator~\cite{del2026electromagnetic}. Yet, advanced signal processing often requires \textit{non-linear} operations. For instance, the universal approximation theorem for neural networks pivotally relies on non-linearity~\cite{hornik1989multilayer,leshno1993multilayer,yarotsky2017error}. Indeed, without non-linearity, any combination of linear transformations collapses to a single equivalent linear operator, precluding the representation of non-linear functions. Identifying and understanding mechanisms for introducing non-linearity in pWDC is an active research topic. At moderate signal levels, wave-based systems typically realize linear input–output mappings, so incorporating non-linearity is not straightforward.

Of course, non-linear circuit blocks (e.g., limiters under strong excitation or saturating gain stages) could enable non-linear processing in the radio-frequency (RF) wave domain, but their non-linearity is usually intended to emerge at relatively high signal intensities. This characteristic is often incompatible with the ultra-low signal levels and energy budgets targeted by pWDC architectures, motivating implementations of non-linear mechanisms outside the RF wave domain. On the one hand, analog electronics can implement some non-linear functions with low latencies and low thresholds; for instance, a ReLU-like function can be implemented by driving a tunable gain element with a DC output signal from an RF detector~\cite{ning2025multilayer}. On the other hand, digital processing can implement arbitrary non-linear functions. Digital non-linearity can then be integrated as non-linear digital backend~\cite{del2020learned} or, with a further digital-to-RF conversion, as non-linear feedback mechanism~\cite{gao2023programmable}. In addition, iterative techniques can update the tunable parameters based on the detected output signals~\cite{tzarouchis2025programmable}.
Requirements for additional conversions between RF and digital domains can involve severe latency and energy efficiency penalties; however, in wave-based sensing there is an inevitable RF-to-digital conversion such that a non-linear digital backend does not require additional conversions between RF and digital domains.

The principles for introducing non-linearity in pWDC discussed thus far all assume that the input data is encoded into the input wavefront. An alternative approach unlocked by \textit{programmable} wave-domain hardware is to encode input data instead into the configuration of the physically tunable parameters. In that case, there exist two qualitatively different mechanisms for non-linearity. On the one hand, the encoding of the input data into the physical scattering properties of the tunable elements can constitute an \textit{encoding non-linearity}. For instance, if the phase of tunable loads is programmable, then the mapping from phase to reflection coefficient is fundamentally non-linear. On the other hand, the mapping from physical scattering properties to output wavefront can constitute a \textit{structural non-linearity}. For instance, if there is significant mutual coupling (MC) between tunable elements, then the mapping from their reflection coefficients to the output wavefront is fundamentally non-linear. 

The idea to encode input data into the scattering characteristics of physically tunable elements underpins backscatter communications which exist at least since the 1940s~\cite{stockman1948communication,brooker2013lev} and are nowadays omnipresent in RFID technology. With the emergence of PMs~\cite{sievenpiper2003two,kamoda201160,clemente20121,cui2014coding}, arrays of many backscatter elements became available. Early works on pWDC with PMs targeted linear matrix-vector multiplication~\cite{del2018leveraging} and high-fidelity reconfigurable signal differentiation~\cite{sol2022meta}. Both of these works embedded the PM inside a rich-scattering enclosure that creates strong MC between the PM elements; it is now well understood that strong inter-element MC boosts the wave-domain flexibility in pWDC hardware~\cite{prod2025mutual,prod2025benefits}. While these early works aimed to implement linear operations,~\cite{del2018leveraging} already explicitly noticed the non-linearity in the mapping from PM configuration to output wavefront. 
A systematic study of this non-linearity based on the coupled-dipole formalism, and corroborated by experiments with a PM in different radio environments, was reported in~\cite{rabault2024tacit}.
An analogous optical system based on a digital micromirror device (DMD, playing the role of the PM) inside an integrating sphere (playing the role of the rich-scattering enclosure) served as the basis for a detailed analysis of the non-linear mapping in~\cite{eliezer2023tunable}. A first application of the non-linear mapping to deep learning was theoretically studied in~\cite{momeni2022electromagnetic}, followed by the first experiment on a wave-based physical neural network (WPNN) whose non-linearity was based on encoding the input data into the configuration of a PM inside a rich-scattering enclosure in~\cite{momeni2023backpropagation}. Subsequently, various theoretical and experimental works in optics reported on WPNNs whose non-linearity originates from encoding input data into structural parameters~\cite{xia2024nonlinear, yildirim2024nonlinear,wanjura2024fully,li2024nonlinear,rahman2025massively,liu2026nonlinear}. In parallel to these works on WPNNs, some works exploit the non-linear mapping from structural parameters to output wavefront to implement matrix inversion~\cite{mohammadi2019inverse,tzarouchis2025programmable,nerini2025analog}, which is a non-linear operation that naturally fits to the MC-aware system model (see Sec.~\ref{sec:system_model}).

Existing works on WPNNs with structural input encoding often do not clearly distinguish between encoding non-linearity and structural non-linearity. In many PM-based experiments, the PM elements are 1-bit-tunable such that the encoding function is affine~\cite{ambiguityaware,reduced_rank}, and thus the encoding non-linearity is weak. In many optical experiments, the phase encoding implies a strong encoding non-linearity while the structural non-linearity is weak or negligible. To the best of our knowledge, a systematic investigation of the interplay between encoding non-linearity and structural non-linearity, as well as the role of the WPNN depth, is missing. The relative contributions of these three mechanisms (encoding non-linearity, structural non-linearity, depth) to the expressivity of WPNNs based on structural input encoding are unclear, not least because in most works the three mechanisms cannot be turned on or off independently to conduct systematic studies.
Clarifying the respective roles of encoding non-linearity, structural non-linearity, and WPNN depth is also practically important because increasing WPNN depth generally increases hardware complexity, footprint, and cost, whereas strong structural non-linearity may be achieved within a compact single-layer WPNN through strong MC.

In this paper, we fill this research gap. 
Our main contributions are summarized as follows:

\begin{enumerate}

    \item We report the first systematic investigation of the expressivity of WPNNs with structural input encoding in terms of the interplay between encoding non-linearity, structural non-linearity, and depth. We examine the influence of these factors, respectively, by considering different encoding mechanisms, different levels of effective MC strength via time gating, and different numbers of WPNN layers. 
    
    \item We report the first physics-consistent model-based analysis of multi-layer WPNNs with structural input encoding, using multiport-network model parameters obtained via a full-wave numerical solver. Prior work was either model-agnostic~\cite{momeni2023backpropagation,xia2024nonlinear,liu2026nonlinear}, or did not account for MC between tunable elements~\cite{yildirim2024nonlinear,li2024nonlinear,rahman2025massively}, or relied on a phenomenological coupled-mode theory with plausible parameters not tied to a concrete implementation~\cite{wanjura2024fully}. 

    \item We compare two multilayer WPNN architectures with structural input encoding. In both cases, we embed the WPNN's learnable parameters within the physically tunable parameters. In the \textit{shared-weights} architecture, the same trainable weights are applied across all WPNN layers. In the \textit{independent-weights} architecture, each layer's weights are independently tunable. The shared-weights architecture allows us to examine the influence of depth without adding tunable parameters.

    \item We propose a compact D-band implementation of a rich-scattering enclosure parametrized by 100 PM elements. Each layer of our WPNNs is based on this PM-parametrized cavity.

    \item We quantify the WPNN's expressivity in terms of the WPNN's performance on a controlled non-linear regression task (approximating filtered noise), which lets us compare expressivity across architectures and choices of non-linearity mechanisms.

\end{enumerate}

\textit{Organization:}
In Sec.~\ref{sec:system_model}, we introduce the multiport-network system model of a PM-parametrized cavity on which our WPNN architectures are based.
In Sec.~\ref{sec:simulation-setup}, we describe the design of our compact D-band PM-parametrized cavity.
In Sec.~\ref{sec_task}, we define the regression task used to probe WPNN expressivity.
In Sec.~\ref{sec:pnn-architectures}, we present the considered WPNN architectures.
In Sec.~\ref{sec:expressivity-factors}, we describe how we systematically control encoding non-linearity, structural non-linearity, and WPNN depth.
In Sec.~\ref{sec:model-training}, we detail the WPNN training.
In Sec.~\ref{sec:results}, we report our results.
In Sec.~\ref{sec:discussion}, we discuss how the different non-linearity mechanisms and depth affect approximation capability under the physical constraints of the system.
In Sec.~\ref{sec:conclusion}, we conclude the paper.

\textit{Notation:}
$\mathbf{A}_{\mathcal{B}\mathcal{C}}$ denotes the block of the matrix $\mathbf{A}$ whose row and column indices belong to the sets $\mathcal{B}$ and $\mathcal{C}$, respectively. $\mathrm{diag}(\mathbf{a})$ denotes the diagonal matrix whose diagonal entries are given by the elements of the vector $\mathbf{a}$. $\mathbf{I}_n$ denotes the $n \times n$ identity matrix. 
$\mathbf{1}_a$ denotes length-$a$ all-ones vector.
For matrix products, we use the convention
$\prod_{m=1}^{M}\mathbf{A}^{(m)}
\triangleq
\mathbf{A}^{(M)}\mathbf{A}^{(M-1)}\cdots\mathbf{A}^{(1)}$.
$\jmath \triangleq \sqrt{-1}$ denotes the imaginary unit.
$\Re\{\cdot\}$ denotes the real part.

\section{System Model of a PM-parametrized Cavity}
\label{sec:system_model}

\begin{figure}[t]
\centering
\includegraphics[width=\columnwidth]{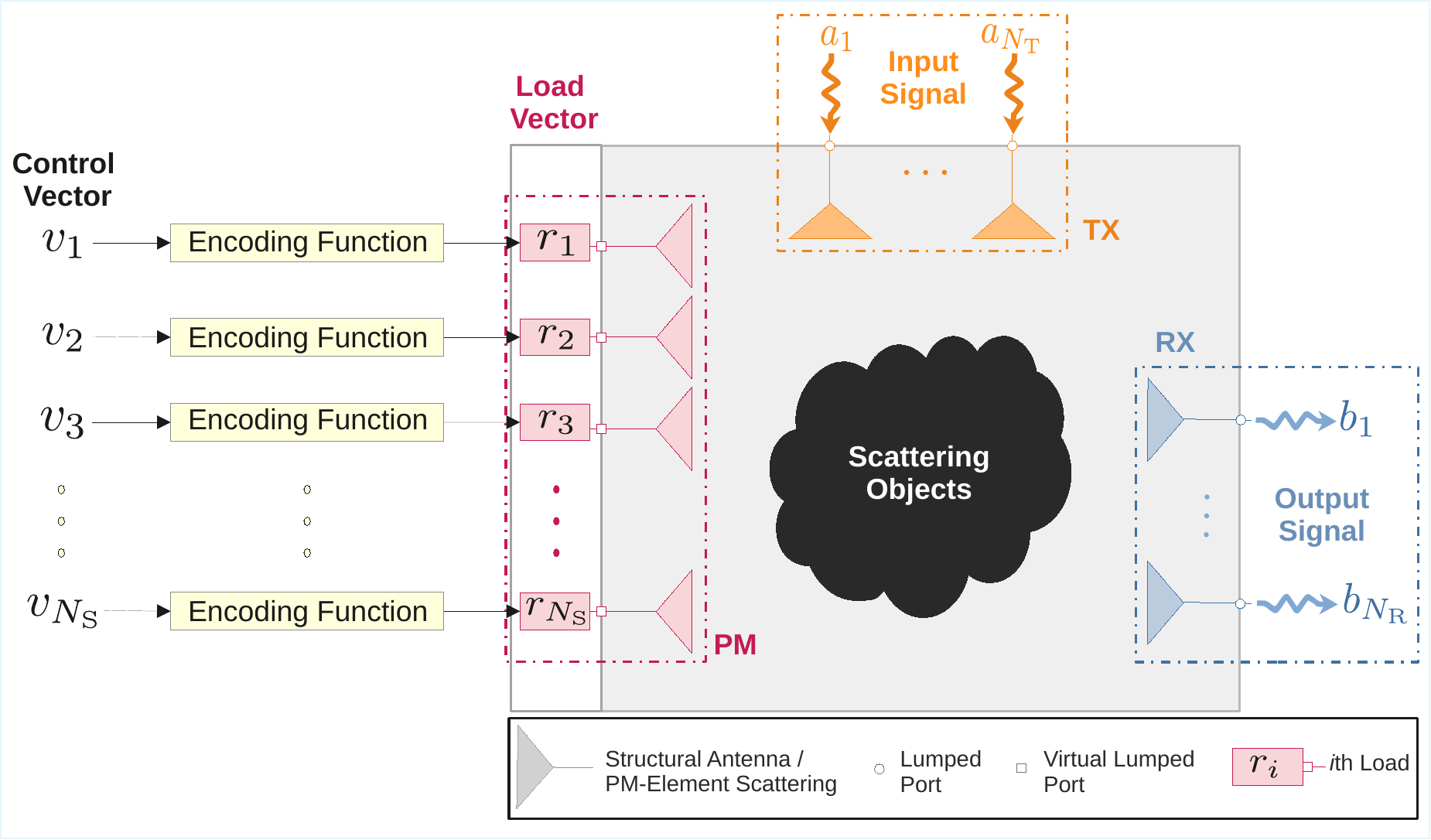}
\caption{System model for a PM-parametrized cavity. The cavity contains $N_\mathrm{T}$ transmitting antennas, $N_\mathrm{R}$ receiving antennas, and a PM with $N_\mathrm{S}$ tunable PM elements. The $i$th PM element is represented as an antenna element whose port is terminated by a tunable load with reflection coefficient $r_i$. The encoding function maps the $i$th entry of the control vector to $r_i$. The static scattering between the antenna and PM ports is described as a linear network.}
\label{fig:system_model}
\end{figure}

In this section, we describe the system model of a PM-parametrized cavity on which we build our WPNN architectures in Sec.~\ref{sec:pnn-architectures}. The concrete implementation and parameter extraction from a full-wave simulation of our PM-parametrized cavity is described in Sec.~\ref{sec:simulation-setup}.

The main assumption of our model is that the PM elements' tunability originates from tunable lumped elements. These tunable lumped elements are thus the only tunable components within the PM-parametrized cavity. We can hence partition the entire system into three entities: (i) $N_\mathrm{A}$ antenna ports via which waves are injected and/or received; (ii) $N_\mathrm{S}$ tunable lumped elements; (iii) an ensemble of all static scattering objects. To be clear, structural scattering of the antennas and PM elements, as well as environmental scattering within the cavity, are captured by (iii). Next, we interpret each tunable lumped element as a ``virtual'' port terminated by a tunable load. Now, as illustrated in Fig.~\ref{fig:system_model}, we can interpret our system as the connection of two multiport networks. Entity (iii) is an $N$-port network characterized by its scattering matrix ${\mathbf{S}}\in\mathbb{C}^{N\times N}$, where $N=N_\mathrm{A}+N_\mathrm{S}$. The ensemble of tunable loads is an $N_\mathrm{S}$-port network, characterized by its diagonal scattering matrix $\mathbf{\Phi}(\mathbf{r})=\mathrm{diag}(\mathbf{r})\in\mathbb{C}^{N_\mathrm{S}\times N_\mathrm{S}}$, where $\mathbf{r}=[r_1, r_2, \dots, r_{N_\mathrm{S}}]\in\mathbb{C}^{N_\mathrm{S}}$ is the load vector whose $i$th entry $r_i$ is the reflection coefficient of the $i$th load. Throughout this paper, we use a reference impedance of $50~\Omega$ at all ports to define all scattering parameters. We further assume that all signal generators and signal detectors are matched to $50~\Omega$.

The two multiport networks are connected via the $N_\mathrm{S}$ ``virtual'' ports, yielding an $N_\mathrm{A}$-port network. We are interested in the end-to-end channel matrix $\mathbf{H}\in\mathbb{C}^{N_\mathrm{R}\times N_\mathrm{T}}$ from one set of $N_\mathrm{T}$ transmitting antennas to a distinct set of $N_\mathrm{R}$ receiving antennas, where $N_\mathrm{T}+N_\mathrm{R}=N_\mathrm{A}$. According to standard multiport network theory (MNT)~\cite{anderson_cascade_1966,ha1981solid}, $\mathbf{H}$ is related to ${\mathbf{S}}$ and $\mathbf{\Phi}(\mathbf{r})$ as follows:
\begin{equation}
\mathbf{H}(\mathbf{r})  = {\mathbf{S}}_\mathcal{RT} + {\mathbf{S}}_\mathcal{RS} \left( \mathbf{I}_{N_\mathrm{S}} - \mathbf{\Phi}(\mathbf{r}) \, {\mathbf{S}}_\mathcal{SS} \right)^{-1} \,\mathbf{\Phi}(\mathbf{r}) \, {\mathbf{S}}_\mathcal{ST} ,
\label{eq1}
\end{equation}
where $\mathcal{T}$, $\mathcal{R}$, and $\mathcal{S}$ denote, respectively, the sets of port indices associated with the transmitting antennas, receiving antennas, and PM elements.
The absence of gain in our system implies that the spectral radius of $\mathbf{\Phi}(\mathbf{r}) \, {\mathbf{S}}_\mathcal{SS}$ is below unity~\cite{del2025physics}, such that (\ref{eq1}) admits the following Neumann-series representation~\cite{zheng2024mutual,wijekoon2024phase,del2025physics}:
\begin{equation}
\begin{split}
\mathbf{H}(\mathbf{r}) & =
  {\mathbf{S}}_\mathcal{RT} 
+ {\mathbf{S}}_\mathcal{RS} 
\left[\sum_{k=0}^{\infty} \left( \mathbf{\Phi}(\mathbf{r}) \, {\mathbf{S}}_\mathcal{SS} \right)^k \right]\, \mathbf{\Phi}(\mathbf{r}) \, {\mathbf{S}}_\mathcal{ST} \\
&=
 \, {\mathbf{S}}_\mathcal{RT} \!
+\! {\mathbf{S}}_\mathcal{RS} \mathbf{\Phi}(\mathbf{r})  {\mathbf{S}}_\mathcal{ST} 
\!+\! {\mathbf{S}}_\mathcal{RS} \mathbf{\Phi}(\mathbf{r})  {\mathbf{S}}_\mathcal{SS} \mathbf{\Phi}(\mathbf{r})  {\mathbf{S}}_\mathcal{ST} 
\!+\! \cdots
\end{split}
\label{eqNeumann}
\end{equation}
In the special case of $\mathbf{S}_\mathcal{SS}=\mathbf{0}$, which implies the absence of MC between the PM elements, the dependence of $\mathbf{H}(\mathbf{r})$ on $\mathbf{\Phi}(\mathbf{r})$ is affine because only the first two terms of the series are non-zero.

In this paper, we assume that each load is continuously and independently tunable. Consequently, the configuration of the $N_\mathrm{S}$ tunable loads is described by a real-valued control vector $\mathbf{v}\in\mathbb{R}^{N_\mathrm{S}}$, where $v_i$ is the control variable applied to the $i$th element. The mapping from $v_i$ to the corresponding physical scattering properties captured by $r_i$ is determined by an elementwise encoding function $c:\mathbb{R}\rightarrow\mathbb{C}$,
\begin{equation}
\label{eq_encoding}
r_i = c(v_i), \qquad i=1,\dots,N_\mathrm{S}.
\end{equation}
Consequently, $\mathbf{r}(\mathbf{v})=[c(v_1),\dots,c(v_{N_\mathrm{S}})]$.

Our system implements a configuration-dependent linear mapping from an input wavefront $\mathbf{a}\in\mathbb{C}^{N_\mathrm{T}}$ injected via the $N_\mathrm{T}$ transmitting antennas to the corresponding output wavefront $\mathbf{b}\in\mathbb{C}^{N_\mathrm{R}}$ exiting the system via the $N_\mathrm{R}$ receiving antennas:
\begin{equation}
    \mathbf{b} = \mathbf{H}(\mathbf{r}(\mathbf{v}))\, \mathbf{a}.
    \label{eq3}
\end{equation}
We emphasize that $\mathbf{b}$ depends linearly on $\mathbf{a}$ but non-linearly on $\mathbf{v}$. We further emphasize that in this work we do \textit{not} encode input data into the input wavefront, but rather into the control vector. Thereby, we can implement a non-linear mapping from input \textit{data} to output \textit{data} based on a wave system that linearly maps input \textit{wavefront} to output \textit{wavefront}.

To summarize, our system model is based on the mapping $\mathbf{v}\rightarrow\mathbf{r}\rightarrow\mathbf{H}$ which involves two generally non-linear functions: \textit{first}, the encoding function in (\ref{eq_encoding}), and, \textit{second}, the MNT function in (\ref{eq1}). The former can constitute an \textit{encoding non-linearity}, while the latter can constitute a \textit{structural non-linearity}. The encoding non-linearity vanishes when $c$ is a linear function. The structural non-linearity vanishes when $\mathbf{S}_\mathcal{SS}=\mathbf{0}$ (and, if one insists on a linear rather than an affine one, then additionally $\mathbf{S}_\mathcal{RT}=\mathbf{0}$ is required). In general, neither $\mathbf{S}_\mathcal{SS}$ nor $\mathbf{S}_\mathcal{RT}$ vanishes.
We describe in Sec.~\ref{sec:expressivity-factors} our choices of encoding function, as well as our systematic control of the encoding non-linearity and structural non-linearity.

\section{Design and Full-Wave Simulation of \\PM-parametrized Cavity}
\label{sec:simulation-setup}

We deliberately work with a PM-parametrized rich-scattering enclosure because rich scattering significantly increases the MC between the PM elements~\cite{rabault2024tacit,reduced_rank}. On the one hand, strong MC boosts the wave-domain flexibility, i.e., the ability to shape $\mathbf{H}$ by controlling $\mathbf{r}$~\cite{prod2025mutual,prod2025benefits}. On the other hand, strong MC results in a strong structural non-linearity. We discuss how we can effectively reduce the MC strength via time gating in post-processing for a benchmark scenario without strong MC in Sec.~\ref{sec:expressivity-factors}.

In addition, we aim to have a compact PM-parametrized cavity, motivating operation at a relatively high frequency. A suitable realization is found in the D-band PM-parametrized wireless network-on-chip (WNoC) studied in~\cite{monochristou_toward_2026}. Here, we re-purpose this architecture that was studied in the distinct context of equalizing wireless channels between cores by optimizing the PM configuration in~\cite{monochristou_toward_2026}. The only difference in our present setup compared to~\cite{monochristou_toward_2026} is that it involves ten rather than seven antennas. For completeness, we briefly describe the entire setup in this section.

\begin{figure}[t]
\centering
\includegraphics[width=\columnwidth]{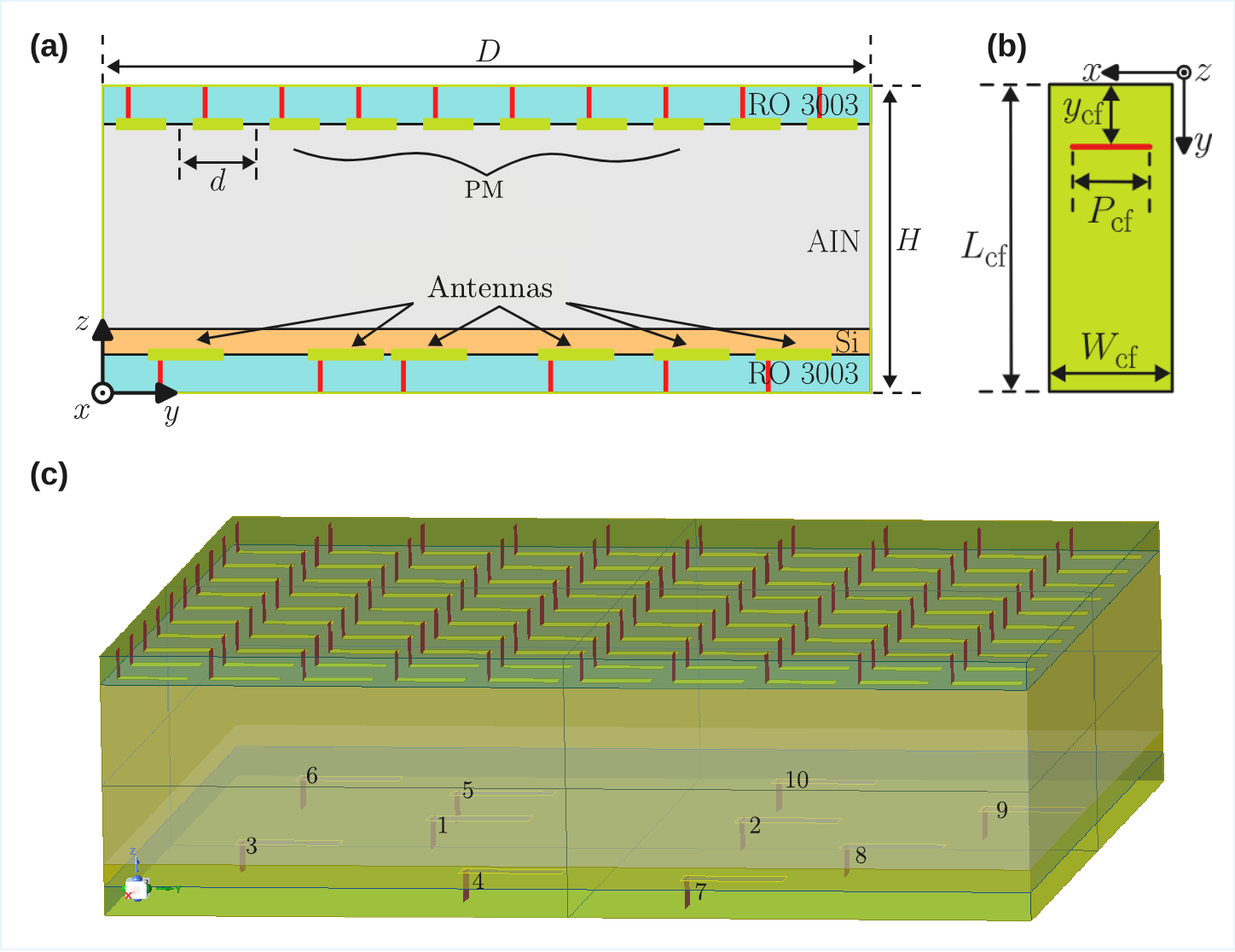}
\caption{(a) Side view of the D-band PM-parametrized rich-scattering enclosure in the WNoC context of~\cite{monochristou_toward_2026}  (see Table~\ref{tb:stackup} for further details). (b) Top view of the microstrip patch element employed as antenna element and as PM element (see Table~\ref{tb:patch_config} for further details). The lumped ports feeding the antenna are depicted in red, and the conductive surfaces in gold. (c) 3D view of the D-band PM-parametrized cavity, with indication of the antenna indices.}
\label{fig2}
\end{figure}

\begin{table}[t]
\centering
\caption{Parameters of the cavity stratification, shown in Fig.~\ref{fig2}. Layer numbering from bottom to top.}
\label{tb:stackup}
\begin{tabulary}{\textwidth}{LCCCC}
\toprule
 & Layer 1 & Layer 2 & Layer 3 & Layer 4 \\
\midrule
Material & Rogers RO3003 & Si & AlN & RO3003 \\
$\varepsilon_\mathrm{r}$ & \num{3} & \num{11.9} & \num{8.8} & \num{3} \\
$h$ & \qty{0.127}{\milli\meter} & \qty{0.1}{\milli\meter} & \qty{0.8}{\milli\meter} & \qty{0.127}{\milli\meter} \\
\bottomrule
\end{tabulary}
\end{table}

\begin{table}[t]
\centering
\caption{Geometrical parameters of the patch elements in Fig.~\ref{fig2}.}
\begin{tabulary}{\textwidth}{LCCCC}
\toprule
Element type & $W_\text{cf}$ & $L_\text{cf}$ & $y_\text{cf}$ & $P_\text{cf}$	\\	
\midrule
Antenna & \qty{0.1}{\milli\meter} & \qty{0.27}{\milli\meter} &  \qty{0}{\milli\meter} &  \qty{0.07}{\milli\meter}
\\
PM element & \qty{0.12}{\milli\meter} & \qty{0.4}{\milli\meter} &  \qty{0.02}{\milli\meter} &  \qty{0.12}{\milli\meter} \\
\bottomrule
\end{tabulary}
\label{tb:patch_config}
\end{table}

At the core of the system displayed in Fig.~\ref{fig2} is a stack of multiple dielectric layers (see details in Table~\ref{tb:stackup}) that represents in a simplified manner the electromagnetically relevant properties of a typical multi-core chip~\cite{timoneda2020engineer,imani2021smart,tapie2023systematic}. This stack is enclosed by conductive walls (representing a thin layer of solder bumps on the bottom and the chip package on the other surfaces~\cite{timoneda2020engineer,imani2021smart,tapie2023systematic}) that we simulate as thin sheets of copper. A $10 \times 10$ array of regularly spaced\footnote{The spacing $d=\qty{0.36}{\milli\meter}$ is half the wavelength inside AlN at \qty{140}{\giga\hertz}.} PM elements is placed at the cavity's top, and an irregular 10-element array of antennas is placed at the cavity's bottom. The antennas and PM elements are both implemented as patch antennas on Rogers substrate (see details in Table~\ref{tb:patch_config}). Each patch contains a lumped port whose size is deeply subwavelength.  

To summarize, we have $N_\mathrm{A}=10$ and $N_\mathrm{S}=100$. As explained in~\cite{tapie2023systematic}, we can extract $\mathbf{S}\in\mathbb{C}^{110\times 110}$ with a single full-wave simulation. We use the commercial ANSYS HFSS software~\cite{noauthor_ansystextsuperscripttextregistered_2023}. We extract $\mathbf{S}$ for 481 regularly spaced frequency points between 110~GHz and 170~GHz. While our operating frequency in this paper is 140~GHz, we need a wideband spectrum for the time-gating procedure described in Sec.~\ref{sec:expressivity-factors} to effectively reduce the MC strength in post-processing. Our system has a transverse dimension of $D=\qty{3.6}{\milli\meter}$ and a height of $H=\qty{1.154}{\milli\meter}$.

\section{WPNN Training Task}
\label{sec_task}

To probe and compare the expressivity of various WPNN architectures in this work, we consider their performance on the regression task of approximating a scalar non-linear function
\begin{equation}
y = f(x),
\end{equation}
where $x \in [0,1]$ and $y \in \mathbb{R}$. 
We generate the target function $f$ by low-pass filtering white Gaussian noise sampled at $n=600$ points uniformly spaced in $[0,1]$. 
The low-pass filter is a 4th-order Butterworth filter with normalized cut-off frequency $f_c$ (relative to the Nyquist frequency) selected from
 $\{0.01, 0.02, \dots, 0.09\}$.
The normalized cut-off frequency serves as a control parameter for the functional complexity of $f$. Increasing the cut-off frequency increases the function’s high-frequency content, resulting in more rapidly varying targets that are typically harder to approximate.
In the limit of a vanishing cut-off frequency, $f$ reduces to a constant. For cut-off frequencies just above zero, the remaining slow variation over the finite interval is approximately affine. For higher cut-off frequencies, $f$ is strongly oscillatory.

To ensure that all target values $y$ lie within the physically reachable output range of the WPNN, we standardize the target signal to zero mean and unit variance and then apply a global scaling factor $g$:
\begin{equation}
y \leftarrow \frac{y-\mu_y}{g\,\sigma_y},
\end{equation}
where $\mu_y$ and $\sigma_y$ denote, respectively, the empirical mean and standard deviation of the generated target samples. 
We choose $g$ based on the distribution of the scalar system readout (defined in Sec.~\ref{sec:pnn-architectures}) evaluated over 1000 random configurations. Specifically, we empirically choose $g=30$ for our system which we found to ensure that all target values $y$ lie comfortably within the distribution of scalar system readouts across 1000 random PM configurations. 

We randomly select $n_\text{train}=200$ of the samples for training, and we use the remaining $n_\text{test}=n-n_\mathrm{train}=400$ unseen samples for evaluating the WPNN's regression performance. The training procedure is discussed in Sec.~\ref{sec:model-training}. The loss metric that we use to quantify the WPNN's regression performance is the normalized mean squared error (NMSE):
\begin{equation}
\mathrm{NMSE} = \frac{\sum_{k=1}^{K} (\hat{y}_k - y_k)^2}{\sum_{k=1}^{K} y_k^2},
\label{eq_NMSE}
\end{equation}
where  $K$ denotes the number of samples in the dataset under consideration, $\hat{y}_k$ is the predicted output and $y_k$ is the target value for the $k$th sample.

\section{WPNN Architectures}
\label{sec:pnn-architectures}

\begin{figure*}[t]
\centering
\includegraphics[width=\linewidth]{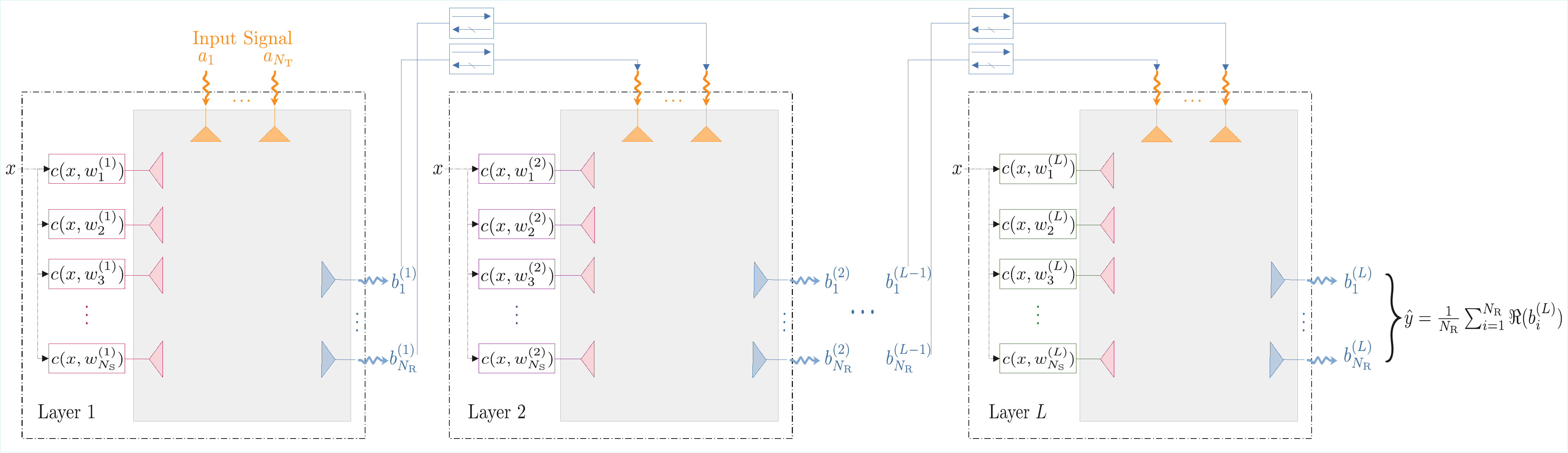}
\caption{\textit{Independent-weights} WPNN architecture. Each layer is based on the PM-parametrized cavity modeled in Fig.~\ref{fig:system_model} whose concrete implementation is shown in Fig.~\ref{fig2}. Using isolators, the output wavefront from one layer is unidirectionally fed into the next layer as input wavefront. Each layer has its own set of learnable weights, which is emphasized visually by different colors of the loads in each layer. The \textit{independent-weights} WPNN architecture specializes to the \textit{shared-weights} WPNN architecture when all layers share the same weights.}
\label{fig:independent_arch}
\end{figure*}

In this section, we describe two multi-layer WPNN architectures. In both cases, each layer is based on the PM-parametrized cavity described in Sec.~\ref{sec:system_model} and Sec.~\ref{sec:simulation-setup}. The key difference between the two architectures is that in one architecture all layers use the same weights (such that we can examine the influence of depth without increasing the number of trainable parameters) while in the other architecture each layer has independent weights. Since the former architecture can be viewed as a special case of the latter architecture, we begin by describing the latter architecture and then explain how it specializes to the former architecture.

As already mentioned, we encode our input data into the control vector rather than into the input wavefront. More precisely, we parametrize the control vector as a function of the input data \textit{and} our trainable weights. Thereby, our trainable weights impact the physical scattering properties of the PM elements; this is in contrast to~\cite{momeni2023backpropagation}, where the PM-parametrized cavity was only used to implement a non-linear data transformation while the trainable weights were in intermediate digital layers.
The $i$th entry of the control vector of the $\ell$th layer is thus defined by a function $q$ that depends on the input data $x$ and the $i$th trainable weight of the $\ell$th layer $w_i^{(\ell)}\in\mathbb{R}$:
\begin{equation}
    v_i^{(\ell)} = q(x,w_i^{(\ell)}).
\end{equation}
We specify $q$ in Sec.~\ref{subsec_EncNL}. For notational ease, we also write $\mathbf{v}^{(\ell)} = q(x,\mathbf{w}^{(\ell)})=
\left[q\!\left(x,w^{(\ell)}_1\right),\; \dots,\; q\!\left(x,w^{(\ell)}_{N_\mathrm{S}}\right)\right]$, where it is understood that $q$ acts elementwise.

For the first layer, we fix the input wavefront to $\mathbf{a}^{(1)}=\mathbf{1}_{N_\mathrm{T}}$. For all other layers, we fix the input wavefront to the output of the previous layer: $\mathbf{a}^{(\ell)}=\mathbf{b}^{(\ell-1)}$. This implies $N_\mathrm{T}=N_\mathrm{R}=5$ (given our PM-parametrized cavity with $N_\mathrm{A}=10$) and unidirectional (and lossless) propagation from the $\ell$th to the $(\ell+1)$th layer. Unidirectional propagation can be imposed with non-reciprocal circuit components such as isolators. The absence of backward inter-layer propagation endows the WPNN's mapping from $\mathbf{a}^{(1)}$ to $\mathbf{b}^{(L)}$, where $L$ is the number of WPNN layers, with a physically meaningful feed-forward factorization into an ordered product of layer-wise end-to-end channel matrices. This factorization defines intermediate wavefronts and thereby renders the notion of ``WPNN depth'' operationally well-defined.

For the $\ell$th layer, the control vector is $\mathbf{v}^{(\ell)} = q(x,\mathbf{w}^{(\ell)})$ and the load vector is $\mathbf{r}^{(\ell)} = c(\mathbf{v}^{(\ell)}) = c(q(x,\mathbf{w}^{(\ell)}))$. Consequently, the end-to-end channel matrix of the $\ell$th layer is $\mathbf{H}^{(\ell)}(c(q(x,\mathbf{w}^{(\ell)})))$. 
The WPNN's scalar readout is based on a simple analog combination of a single coherent quadrature (I-channel) across the $N_\mathrm{R}$ receive ports. Specifically, we define the WPNN's scalar readout in terms of $\mathbf{b}^{(L)}$ as the mean of the real parts of the entries of $\mathbf{b}^{(L)}$:
\begin{equation}
\hat{y}(x,\mathbf{W}) \triangleq \frac{1}{N_\mathrm{R}} \sum_{i=1}^{N_\mathrm{R}} 
\Re\!\left\{b_i^{(L)}(x,\mathbf{W})\right\},
\label{eq:readout}
\end{equation}
where $\mathbf{W} = \left[\mathbf{w}^{(1)}, \dots , \mathbf{w}^{(L)} \right] \in \mathbb{R}^{N_\mathrm{S} \times L}$ stacks the $L$ weight vectors.

So far, we have described our \textit{independent-weights} WPNN architecture which is depicted in Fig.~\ref{fig:independent_arch}. This architecture specializes to the \textit{shared-weights} architecture if we impose $\mathbf{w}^{(i)} = \mathbf{w}^{(1)} \ \forall \ 1<i\leq L$.

\section{Systematic Control of \\Encoding Non-Linearity, Structural Non-Linearity, and Depth} 
\label{sec:expressivity-factors}

As mentioned in the introduction, our goal is to systematically examine the influence of the interplay of (i) encoding non-linearity, (ii) structural non-linearity, and (iii) WPNN depth on the WPNN's expressivity. 
By independently varying these three factors, we can isolate and analyze their respective contributions to the WPNN's expressivity.
In this section, we describe how we systematically control each of these mechanisms.

\subsection{Encoding Non-Linearity}
\label{subsec_EncNL}

The encoding non-linearity is governed by the encoding function defined in (\ref{eq_encoding}). In this theoretical work, we are not constrained by the encoding function associated with any particular lumped tunable element (e.g., varactor diode). Consequently, we can freely choose different encoding functions. We thus consider the following two encoding functions; the first has a strong encoding non-linearity, while the second has no encoding non-linearity.

\subsubsection{Phase Encoding} 
\label{subsubsec_PhaseEncoding}

We assume that the magnitudes of the loads' reflection coefficients are fixed to unity and that the phases of the loads' reflection coefficients can be tuned continuously without constraint. Similar phase-encoding assumptions are commonly made in theoretical literature on wireless communications with reconfigurable intelligent surfaces, as well as in works on optical WPNNs~\cite{yildirim2024nonlinear,li2024nonlinear,rahman2025massively}. In summary, in the case of phase encoding we work with
\begin{subequations}
\label{eq_phase_encoding}
    \begin{equation}
        v_i^{(\ell)}  = q_\mathrm{P}(x,w_i^{(\ell)}) = x+w_i^{(\ell)},
    \end{equation}
    \begin{equation}
        r_i^{(\ell)} = c_\mathrm{P}(v_i^{(\ell)}) = \mathrm{e}^{\jmath 2 \pi v_i^{(\ell)}}.
    \end{equation}
\end{subequations}
A physical embodiment of a tunable load enabling such phase encoding is a length-tunable, open-circuited, lossless delay line.

\subsubsection{Linear Encoding}
\label{subsubsec_LinEncoding}

To consider a case without encoding non-linearity, we assume that the reflection coefficient of the $i$th load in the $\ell$th layer is the product of the input data and the $i$th weight in the $\ell$th layer. This choice eliminates by construction the encoding non-linearity. To ensure that the loads remain passive, i.e., $|r_i^{(\ell)}| \leq 1 \ \forall \ i,\ell$, it is sufficient to clip $w_i^{(\ell)}$ since $x \in [0,1]$ by construction (see Sec.~\ref{sec_task}).
In summary, in the case of linear encoding we work with
\begin{subequations}
\label{eq_linear_encoding}
    \begin{equation}
        v_i^{(\ell)}  = q_\mathrm{L}(x,w_i^{(\ell)}) = x\,\tilde{w}_i^{(\ell)},
    \end{equation}
    \begin{equation}
        r_i^{(\ell)} = c_\mathrm{L}(v_i^{(\ell)}) =  v_i^{(\ell)},
    \end{equation}
\end{subequations}
where $\tilde{w}_i^{(\ell)} = \min\!\big(\max(w_i^{(\ell)},0),1\big)$ denotes a clipped version of $w_i^{(\ell)}$.

\subsection{Structural Non-Linearity}
\label{subsec_timegating}

\begin{figure}[t]
\centering
\includegraphics[width=\linewidth]{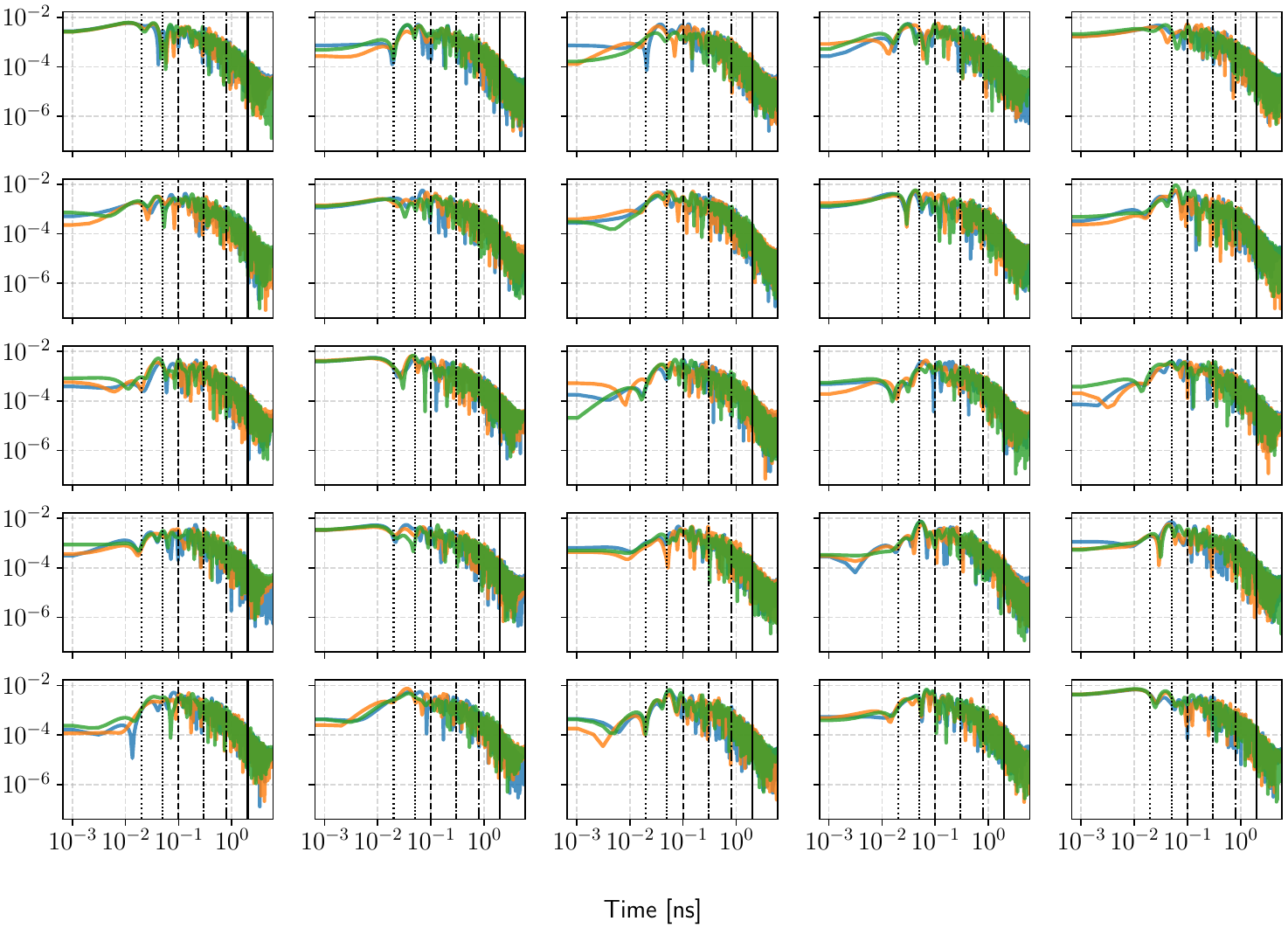}
\caption{End-to-end channel impulse responses of our PM-parametrized cavity (with $N_\mathrm{T}=N_\mathrm{R}=5$) depicted in Fig.~\ref{fig2} for three random PM configurations (color-coded). The six vertical lines indicate the six considered finite time-gating truncation times: 0.02~ns, 0.05~ns, 0.1~ns, 0.3~ns, 0.8~ns, and 2.0~ns.}
\label{fig4}
\end{figure}

The structural non-linearity is governed by the MC between the PM elements,  mediated by the $\mathbf{S}_\mathcal{SS}$ term in (\ref{eq1}). It would be impractical to re-design and re-simulate the entire PM-parametrized cavity for each considered MC strength. Instead, following~\cite{prod2025benefits}, we adopt a convenient post-processing technique that consists of adjusting the system's \textit{effective} MC strength via time gating. MC can be understood in terms of a picture of multi-bounce paths~\cite{rabault2024tacit}. Longer paths are typically associated with longer dwell times within the system and more scattering events along their trajectories. The more often a path interacts with different tunable lumped elements, the more it intertwines their effects on the end-to-end channel matrix. In other words, strong MC is associated with long dwell times.

Time gating is a standard post-processing technique for suppressing scattered waves in imaging applications~\cite{7362172,barolle2021manifestation,9926204} and antenna characterization~\cite{1296154,timegating_XLIM}. As in~\cite{prod2025benefits}, we apply time gating here to suppress the contributions of paths associated with delay times beyond $\tau$, thereby  reducing the system's \textit{effective} MC strength. In other words, the truncation time $\tau$ is a proxy for the \textit{effective} MC strength and provides us with continuous control in post-processing over the system's MC strength by selecting $\tau$. An upper bound on the MC strength is of course the system's inherent MC strength that is reached in the limit of $\tau\rightarrow \infty$ (i.e., when no time gating is applied).

Specifically, we time-gate the end-to-end channel matrix $\mathbf{H}$. To that end we inverse-Fourier transform the wideband spectrum ($110-170$~GHz) of each entry of $\mathbf{H}$ to the time domain, we truncate the time-domain signal at delay time $\tau$, we Fourier transform the truncated time-domain signal to the frequency domain, and we select its frequency point at 140~GHz.
We consider seven truncation times: 0.02~ns (which corresponds roughly to the first prominent peak in the raw end-to-end channel impulse responses), 0.05~ns, 0.1~ns, 0.3~ns, 0.8~ns, 2.0~ns, and $\infty$ (i.e., no time gating). We indicate six finite truncation times in Fig.~\ref{fig4} where we plot the 25 end-to-end channel impulse responses for three random PM configurations.

To summarize, a lower truncation time $\tau$ corresponds to a weaker structural non-linearity. Thus, $\tau$ is a convenient control knob to adjust the strength of the structural non-linearity. 
If $\tau$ is chosen smaller than the delay time of the shortest path from transmitters to receivers that encountered the PM at least once, then the resulting time-gated end-to-end channel matrix is dominated by PM-independent components and the WPNN output becomes (approximately) independent of the input data and the trainable weights.

\subsection{Depth}

The WPNN depth is easily controlled by choosing the number of layers $L$. In the independent-weights architecture, increasing $L$ simultaneously increases the number of trainable weights; in contrast, in the shared-weights architecture, the number of trainable weights does not depend on $L$.

\section{WPNN Training}
\label{sec:model-training}

Because we have fully differentiable forward models of our WPNNs (see Sec.~\ref{sec:system_model} and Sec.~\ref{sec:pnn-architectures}) with knowledge of the fixed model parameters (i.e., the relevant entries of $\mathbf{S}$ determined via a full-wave numerical simulation in Sec.~\ref{sec:simulation-setup}) and the feasible set of the trainable parameters (i.e., $\mathbf{W}$), we can conveniently train our WPNNs end-to-end using standard gradient-based optimization. Concretely, we implement our WPNNs in the automatic-differentiation framework PyTorch such that we are not required to provide analytic derivative expressions. 
We optimize the trainable parameters by minimizing the NMSE computed over a mini-batch of 100 training samples using the Adam optimizer with learning rate $10^{-3}$ for 1000 iterations.
In the case of linear encoding (see Sec.~\ref{subsubsec_LinEncoding}), we enforce the constraint $0\le w_i^{(\ell)}\le 1$ by clipping the entries of $\mathbf{W}$ after each update. 
In the case of time gating with a finite $\tau$, we implement the required inverse Fourier transforms and Fourier transforms as part of our forward model.

Our WPNN training can be characterized as a model-based in-silico training strategy, i.e., we optimize the trainable weights $\mathbf{W}$ on a conventional digital processor by backpropagating through a differentiable forward model of the WPNN. Intense contemporary research efforts are dedicated to alternative training techniques for physical neural networks to mitigate concerns over model-reality mismatch or to operate in a model-agnostic manner~\cite{momeni2025training}. While the feasibility of obtaining an accurate forward model is an important concern for many physical neural networks, we are confident that it is possible to obtain an accurate forward model for an experimental implementation of the particular WPNN architecture based on PM-parametrized cavities studied in this work. Indeed, various recent works demonstrated the experimental estimation of a set of proxy parameters for the system model described in Sec.~\ref{sec:system_model} to accurately model the mapping from $\mathbf{v}$ to $\mathbf{H}$ in various PM-parametrized cavities~\cite{sol2024experimentally,del2025experimental,ambiguityaware,reduced_rank}. These works give us confidence that model-based in-silico training  can  be applied to experimental prototypes of the WPNNs considered in this work.

\begin{figure*}[t]
\centering
\includegraphics[width=0.7\textwidth]{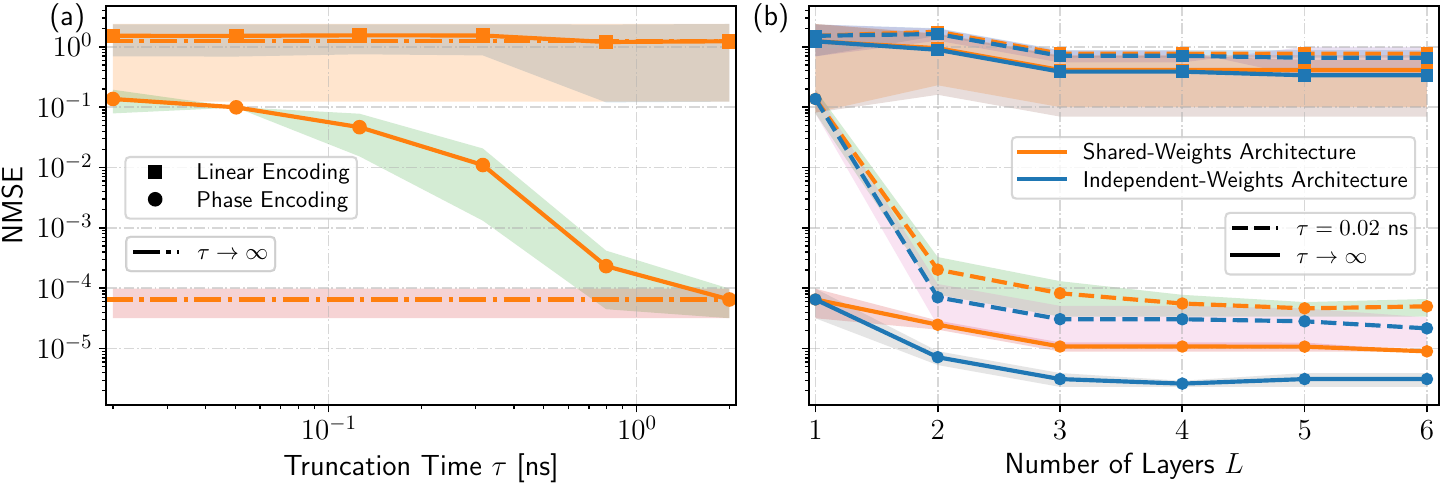}
\caption{Non-linear regression performance on 50 target functions obtained with $f_c=0.02$; the curves display the median NMSE across the 50 target functions, and the shades indicate the corresponding standard deviation. 
(a) Performance dependence on the strength of the structural non-linearity in the single-layer case, for both choices of encoding function (phase encoding and linear encoding); the two WPNN architectures (independent-weights and shared-weights) are identical in the single-layer case. As described in Sec.~\ref{subsec_timegating}, we control the strength of the structural non-linearity via the truncation time $\tau$ used for time gating that acts as a proxy for the effective MC strength. The horizontal lines correspond to the limit $\tau\rightarrow\infty$ (i.e., no time gating). 
(b) Performance dependence on the WPNN depth, for both WPNN architectures (independent-weights and shared-weights), both choices of encoding function (phase encoding and linear encoding), and two choices of effective MC strength ($\tau=0.02$~ns and $\tau\rightarrow\infty$). 
}
\label{Fig5}
\end{figure*}

\begin{figure*}[t]
\centering
\includegraphics[width=\textwidth]{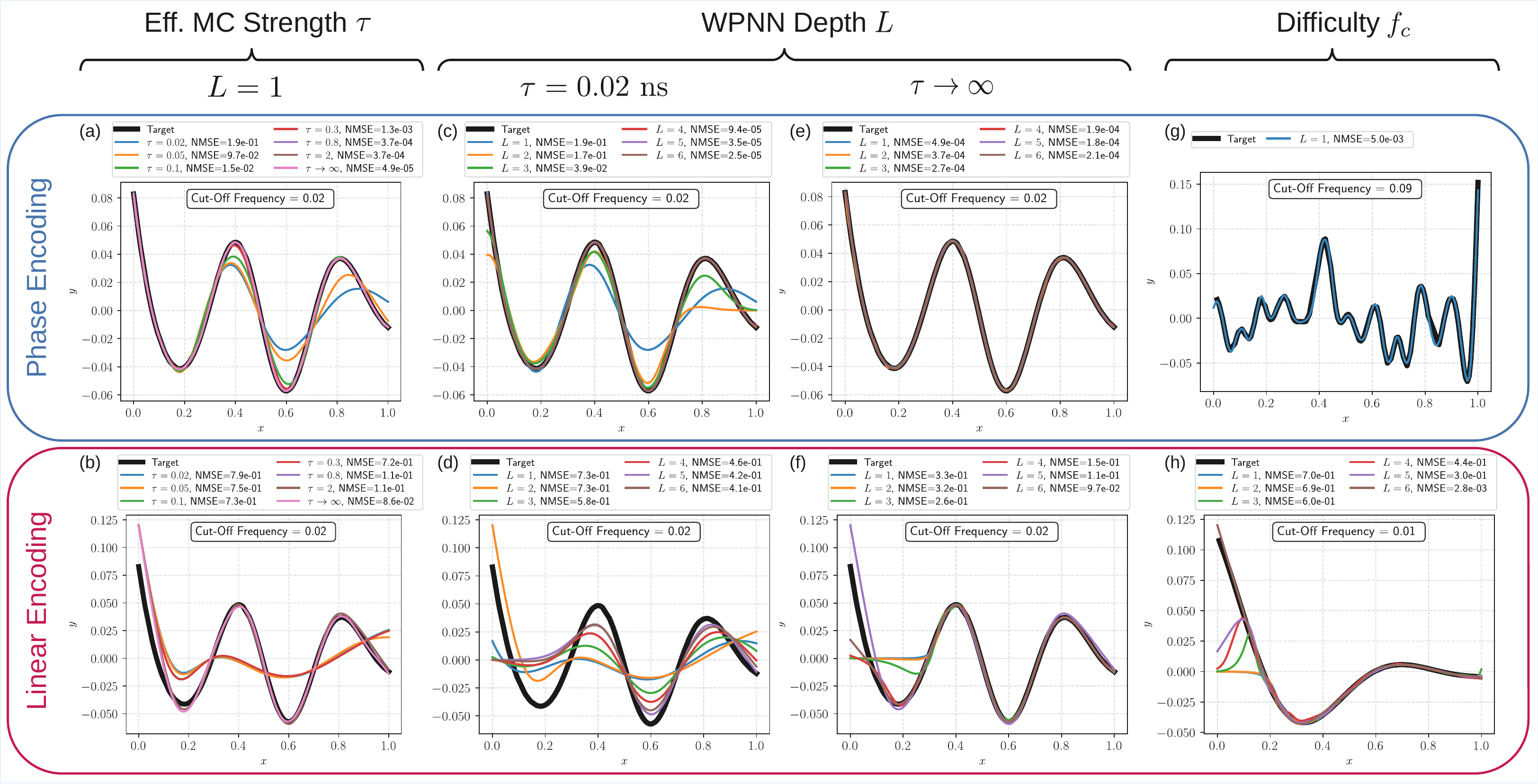}
\caption{Selected examples of target functions and independent-weights WPNN approximations thereof. (a,b) correspond to Fig.~\ref{Fig5}a. (c,d,e,f) correspond to Fig.~\ref{Fig5}b. (g,h) correspond to Fig.~\ref{Fig7}. In each case, the corresponding NMSE is indicated.}
\label{Fig6}
\end{figure*}

\section{Results}
\label{sec:results}

In this section, we systematically study the expressivity of our WPNN architectures. We quantify the WPNN's non-linear regression performance in terms of the median NMSE achieved for 50 target functions with a given cut-off frequency $f_c$ that determines the level of difficulty. 

We begin by examining the influence of the structural non-linearity in the single-layer case in Fig.~\ref{Fig5}a, considering a medium difficulty level of $f_c=0.02$. As explained in Sec.~\ref{subsec_timegating}, we vary the time-gating truncation time $\tau$ to control the effective MC strength, which in turn governs the strength of the structural non-linearity. In this single-layer case, the independent-weights and shared-weights architectures are identical. For phase encoding, we observe that the performance improves monotonically as we increase $\tau$. At the lowest considered value of $\tau$, the performance is not satisfactory. A corresponding example is visualized in Fig.~\ref{Fig6}a. At the largest value of $\tau$ (i.e., without time gating), the NMSE is more than three orders of magnitude lower and the corresponding example shown in Fig.~\ref{Fig6}a is flawless. 
Altogether, phase encoding combined with strong structural non-linearity thus achieves good performance in a shallow WPNN. Limiting the number of WPNN layers is attractive for reducing physical footprint and cost. Strong structural non-linearity is thus a viable route to working with a shallow WPNN based on phase encoding.
In contrast, for linear encoding, the NMSE is seen in Fig.~\ref{Fig5}a to be much higher and to only marginally improve upon increasing $\tau$. The illustrative examples in Fig.~\ref{Fig6}b confirm that the WPNN's approximation of the target function is imperfect in these cases. However, careful inspection of Fig.~\ref{Fig6}b reveals that the WPNN's approximation is in fact good for larger values of $x$ while it is poor for lower values of $x$. This observation can be understood from (\ref{eq_linear_encoding}). Indeed, for small $x$, the reflection coefficients $r_i = x\tilde{w}_i$ are necessarily small for all $i$, regardless of $\tilde{w}_i$. Consequently, the PM-dependent contribution to the WPNN output is weak, and the scalar readout depends only weakly on $x$. The PM-independent contribution to the WPNN's scalar readout dominates but is not controllable. This appears to be a fundamental shortcoming of the considered linear encoding that could be mitigated with affine encoding.

Next, we examine the influence of the number of layers $L$ in Fig.~\ref{Fig5}b, considering again a medium difficulty level of $f_c=0.02$. We compare phase encoding and linear encoding, $\tau=0.02$~ns and $\tau\rightarrow \infty$, as well as shared-weights and independent-weights architectures. In line with our observation in Fig.~\ref{Fig5}a, we see that phase encoding without time gating performs very well already with $L=1$. With more layers, the performance improves by almost an order of magnitude for the shared-weights architecture and by more than an order of magnitude for the independent-weights architecture. The former evidences some benefits of depth; the latter evidences  benefits from depth combined with additional trainable parameters. Upon inspection of the examples in Fig.~\ref{Fig6}e, the benefits for multiple layers are not visually apparent. In contrast, with $\tau=0.02$~ns, the WPNN depth can  compensate the absence of significant structural non-linearity. A particularly sharp drop in NMSE by roughly three orders of magnitude is seen in Fig.~\ref{Fig5}b as we go from one layer to two layers in the case of $\tau=0.02$~ns. Additional layers yield marginal further reductions in NMSE. The difference between shared-weights architecture and independent-weights architecture is small, indicating that the main benefit of WPNN depth does not originate from introducing additional trainable parameters. The examples in Fig.~\ref{Fig6}c corroborate these observations.
For linear encoding, the NMSE is always very high, for the same reason already discussed before, namely that the considered linear encoding cannot achieve good approximation fidelity for small $x$. This is again clearly visible in the examples in Fig.~\ref{Fig6}f.

\begin{figure}
\centering
\includegraphics[width=\columnwidth]{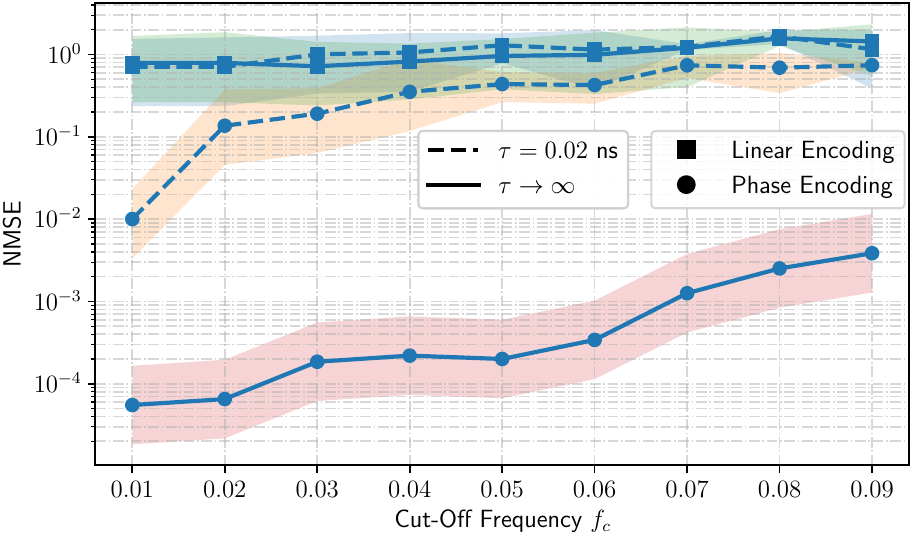}
\caption{Dependence of NMSE on the difficulty level determined by the cut-off frequency $f_c$, for phase encoding and linear encoding as well as for $\tau=0.02$~ns and $\tau\rightarrow\infty$.}
\label{Fig7}
\end{figure}

Finally, we systematically examine the dependence of the NMSE on the level of difficulty controlled by the cut-off frequency $f_c$. For the case of phase encoding, we display an example with high difficulty level ($f_c=0.09$) in Fig.~\ref{Fig6}g that is mastered by a single-layer WPNN with phase encoding and strong structural non-linearity. We further systematically investigate the scaling of the NMSE with $f_c$ in Fig.~\ref{Fig7}. We see that the NMSE monotonically increases with $f_c$ but remains reasonably low even for the highest considered difficulty level. For the case of linear encoding, we display an example with low difficulty level ($f_c=0.01$) in Fig.~\ref{Fig6}h. The approximation is good for $x>0.3$ in all cases but poor for smaller values of $x$ except for the case with $L=6$. Substantial depth can thus  alleviate the shortcoming of the considered linear encoding, at least for a low-difficulty target function.

\section{Discussion}
\label{sec:discussion}

In this section, we interpret the results from Sec.~\ref{sec:results} using our system model (Sec.~\ref{sec:system_model} and Sec.~\ref{sec:pnn-architectures}) in instructive limiting cases. Specifically, we consider (i) single-layer WPNNs with strong mutual coupling (MC) and (ii) multi-layer WPNNs without MC, for phase encoding (Sec.~\ref{sec_discuss_PhaseEncoding}) and linear encoding (Sec.~\ref{sec_discuss_LinearEncoding}). In the phase-encoding case, the readout can be written as an (infinite or truncated) Fourier series in the scalar input $x$; in the linear-encoding case, it becomes an (infinite or truncated) polynomial in $x$. While both polynomial expansions and trigonometric expansions can approximate broad classes of functions when their coefficients are freely tunable~\cite{fejer1903untersuchungen,Weierstrass1885a}, in our WPNN architectures the coefficients are \emph{not} free: they are constrained by the physics through $\mathbf{S}$ and by the parametrization in terms of $\mathbf{W}$. In the strong-MC single-layer case, higher-order terms exist but their magnitudes typically decay with order; in the no-MC multi-layer case, the series is explicitly truncated, with the maximal order being determined by the depth $L$. These constraints explain why encoding choice, MC strength, and depth play distinct and partially interchangeable roles in the observed expressivity.

Phase encoding induces Fourier terms (i.e., $\{1,\mathrm{e}^{\jmath 2\pi x},\mathrm{e}^{\jmath 4\pi x},\dots\}$), which form an orthogonal set on $[0,1]$. By contrast, with linear encoding the induced basis is a power series in $x$ (i.e., $\{1,x,x^2,\dots\}$), whose terms are highly correlated on $[0,1]$. The orthogonality of the Fourier terms suggests that phase encoding can provide a more representationally efficient scaffold for oscillatory target functions, especially with a limited number of significant series terms and with the series coefficients being constrained by the physical parametrization rather than being freely tunable. 
However, given the constrained control over the coefficients of these series in our WPNNs, differences in basis orthogonality are likely not the dominant effect.

\subsection{Phase Encoding}
\label{sec_discuss_PhaseEncoding}

\subsubsection{Single-Layer WPNN with Phase Encoding and Strong MC}
\label{sec_discuss_A}

Based on (\ref{eq_phase_encoding}), we can write $\mathbf{\Phi}(x,\mathbf{w}) = \mathrm{e}^{\jmath 2 \pi x} \,\mathbf{\Psi}(\mathbf{w})$, where $\mathbf{\Psi}(\mathbf{w}) = \mathrm{diag}([\mathrm{e}^{\jmath 2 \pi w_1},\dots,\mathrm{e}^{\jmath 2 \pi w_{N_\mathrm{S}}}])$ and we drop the superscript $^{(\ell)}$ indicating the layer index $\ell$ since we are in the single-layer case. Substituting $\mathbf{\Phi}(x,\mathbf{w}) = \mathrm{e}^{\jmath 2 \pi x} \,\mathbf{\Psi}(\mathbf{w})$ into (\ref{eqNeumann}), we obtain

\begin{equation}
\begin{split}
\mathbf{H}(x,\mathbf{w})   =\, \,
  &{\mathbf{S}}_\mathcal{RT} 
+ \\ &{\mathbf{S}}_\mathcal{RS} \!
\left[\sum_{k=0}^{\infty} \left(  \mathrm{e}^{\jmath 2 \pi x} \mathbf{\Psi}(\mathbf{w})  {\mathbf{S}}_\mathcal{SS} \right)^k \right] \mathrm{e}^{\jmath 2 \pi x} \mathbf{\Psi}(\mathbf{w})  {\mathbf{S}}_\mathcal{ST} \\
= &\sum_{\bar{k}=0}^{\infty} \left(\mathbf{C}_{\bar{k}}(\mathbf{w}) \, \mathrm{e}^{\jmath 2 \pi \bar{k} x}\right),
\end{split}
\label{eqNeumann12}
\end{equation}
where
\begin{equation}
\mathbf{C}_{\bar{k}}(\mathbf{w}) \triangleq
\begin{cases}
\mathbf{S}_{\mathcal{RT}}, & \bar{k} = 0,\\[4pt]
\mathbf{S}_{\mathcal{RS}}
\left(\mathbf{\Psi}(\mathbf{w})\,\mathbf{S}_{\mathcal{SS}} \right)^{\bar{k}-1}
\,\mathbf{\Psi}(\mathbf{w})\,\mathbf{S}_{\mathcal{ST}}, & \bar{k}>0.
\end{cases}
\label{eq:Ck_cases}
\end{equation}
Substituting (\ref{eqNeumann12}) into (\ref{eq:readout}), we obtain
\begin{equation}
\begin{split}
\hat{y}(x,\mathbf{w})
=\Re\!\left\{\sum_{\bar{k}=0}^{\infty} \alpha_{\bar{k}}(\mathbf{w}) \, \mathrm{e}^{\jmath 2\pi \bar{k} x}\right\}.
\end{split}
\label{eq:yhat_onesTb}
\end{equation}
where 
\begin{equation}
    \alpha_{\bar{k}}(\mathbf{w}) \triangleq
\left(\frac{1}{N_\mathrm{R}}\mathbf{1}_{N_\mathrm{R}}\right)^\top
\mathbf{C}_{\bar{k}}(\mathbf{w})\,\mathbf{1}_{N_\mathrm{T}}.
\label{alpha_k}
\end{equation}
Consequently, our scalar readout signal $\hat{y}(x,\mathbf{w})$ is the real part of a complex Fourier series in the scalar input $x$. 
Stronger MC typically leads to a slower decay of the Neumann series such that more higher-order Fourier terms contribute significantly. In addition, we expect that stronger MC and larger $N_\mathrm{S}$ improve the ability to shape the Fourier coefficients.

\subsubsection{Multi-Layer WPNN with Phase Encoding and No MC}
\label{sec_discuss_B}

Based on (\ref{eq_phase_encoding}) and under the assumption of $\mathbf{S}_\mathcal{SS}=\mathbf{0}$ (which implies the absence of MC), (\ref{eq1}) collapses to
\begin{equation}
    \mathbf{H}^{(\ell)}(x,\mathbf{w}^{(\ell)})=
    \mathbf{S}_\mathcal{RT}+\mathrm{e}^{\jmath 2 \pi x}\,\mathbf{B}^{(\ell)}(\mathbf{w}^{(\ell)}),
\end{equation}
where $\mathbf{B}^{(\ell)}\!\left(\mathbf{w}^{(\ell)}\right)
\triangleq
\mathbf{S}_\mathcal{RS}\,\mathbf{\Psi}(\mathbf{w}^{(\ell)})\,\mathbf{S}_\mathcal{ST}$.
It follows that
\begin{equation}
\begin{split}
    \mathbf{b}^{(L)} (x,\mathbf{W})
=\left[
\prod_{m=1}^{L} \left( \mathbf{S}_\mathcal{RT} + \mathrm{e}^{\jmath 2 \pi x} \, \mathbf{B}^{(m)}(\mathbf{w}^{(m)}) \right)
\right]\mathbf{1}_{N_\mathrm{T}}.
\end{split}
\label{eq:b_ell_prod_forward}
\end{equation}
Since each factor in \eqref{eq:b_ell_prod_forward} is affine in
$\mathrm{e}^{\jmath 2\pi x}$, the ordered product expands into a finite sum
$\sum_{k=0}^{L} \mathbf{D}_k(\mathbf{W}) \, \mathrm{e}^{\jmath 2\pi k x}$,
where the coefficient matrices $\mathbf{D}_k(\mathbf{W})$ are given by
\begin{equation}
\mathbf{D}_{k}(\mathbf{W})
\triangleq \!\!\!\!\!\!\!
\sum_{1\le i_{1}<\cdots<i_{k}\le L}\!\!\!\!\!\!\!
\mathbf{S}_\mathcal{RT}^{\,L-i_{k}}\,
\mathbf{B}^{(i_{k})}\,
\mathbf{S}_\mathcal{RT}^{\,i_{k}-i_{k-1}-1}\,
\cdots\,
\mathbf{B}^{(i_{1})}\,
\mathbf{S}_\mathcal{RT}^{\,i_{1}-1},
\label{eq:Dk_def}
\end{equation}
with $\mathbf{D}_0(\mathbf{W})=\mathbf{S}_\mathcal{RT}^{L}$; recall that our WPNN architectures assume $N_\mathrm{R}=N_\mathrm{T}$ such that $\mathbf{S}_\mathcal{RT}$ is a square matrix and its powers are well-defined.
Consequently, the WPNN's scalar readout defined in \eqref{eq:readout} becomes  a truncated complex Fourier series in $x$, whose coefficients are parametrized by $\mathbf{W}$:
\begin{equation}
\hat{y}(x,\mathbf{W})
=
\Re\!\left\{
\sum_{k=0}^{L} \beta_k(\mathbf{W})\,\mathrm{e}^{\jmath 2\pi k x}
\right\},
\label{eq:yhat_truncFS}
\end{equation}
where
\begin{equation}
\beta_{k}(\mathbf{W})
\triangleq
\left(\frac{1}{N_\mathrm{R}}\,\mathbf{1}_{N_\mathrm{R}}\right)^{\top}
\mathbf{D}_{k}(\mathbf{W}) \, \mathbf{1}_{N_\mathrm{T}}.
\label{eq:beta_def}
\end{equation}

We see in (\ref{eq:yhat_truncFS}) that even in the absence of MC, depth $L$ generates Fourier terms up to order $L$ in $x$. In contrast to the strong-MC single-layer case in Sec.~\ref{sec_discuss_A}, the number of available Fourier terms is now explicitly capped by the WPNN's depth. As in Sec.~\ref{sec_discuss_A}, the Fourier coefficients cannot be chosen freely but are parametrized by $\mathbf{W}$.

In the special case $\mathbf{S}_\mathcal{RT}=\mathbf{0}$, the expression for $\mathbf{D}_k$ simplifies to $\mathbf{D}_k=\mathbf{0} \ \forall \ k<L$ and $\mathbf{D}_L = \prod_{m=1}^{L} \mathbf{B}^{(m)}$. Consequently, the expressivity is severely limited in this case, highlighting that a non-vanishing $\mathbf{S}_\mathcal{RT}$ is desirable in the case of a multi-layer WPNN with phase encoding and no MC.

\subsection{Linear Encoding}
\label{sec_discuss_LinearEncoding}

\subsubsection{Single-Layer WPNN with Linear Encoding and Strong MC}
\label{sec_discuss_Lin_single_strongMC}

Based on \eqref{eq_linear_encoding}, we can write $\mathbf{\Phi}(x,\mathbf{w})= x\, \mathbf{\Omega}(\mathbf{w})$, where $ \mathbf{\Omega}(\mathbf{w}) = \mathrm{diag}(\tilde{\mathbf{w}})$ and $\tilde{\mathbf{w}}$ is the clipped version of $\mathbf{w}$, as discussed in Sec.~\ref{subsubsec_LinEncoding}. We drop the superscript $^{(\ell)}$ indicating the layer index $\ell$ since we are in the single-layer case. Substituting $\mathbf{\Phi}(x,\mathbf{w})= x\, \mathbf{\Omega}(\mathbf{w})$ into (\ref{eqNeumann}), we obtain
\begin{equation}
\begin{split}
\mathbf{H}(x,\mathbf{w})
&= \mathbf{S}_{\mathcal{RT}}
+\mathbf{S}_{\mathcal{RS}}
\left[\sum_{k=0}^{\infty}\left(x\,\mathbf{\Omega}(\mathbf{w})\,\mathbf{S}_{\mathcal{SS}}\right)^k\right]
x\,\mathbf{\Omega}(\mathbf{w})\,\mathbf{S}_{\mathcal{ST}}\\
&=\sum_{\bar{k}=0}^{\infty}\left(\mathbf{E}_{\bar{k}}(\mathbf{w})\,x^{\bar{k}}\right),
\end{split}
\label{eq:Lin_single_power_series_H}
\end{equation}
where
\begin{equation}
\mathbf{E}_{\bar{k}}(\mathbf{w}) \triangleq
\begin{cases}
\mathbf{S}_{\mathcal{RT}}, & \bar{k}= 0,\\[4pt]
\mathbf{S}_{\mathcal{RS}}
\left(\mathbf{\Omega}(\mathbf{w})\,\mathbf{S}_{\mathcal{SS}}\right)^{\bar{k}-1}
\mathbf{\Omega}(\mathbf{w})\,\mathbf{S}_{\mathcal{ST}}, & \bar{k}>0.
\end{cases}
\label{eq:Ek_cases}
\end{equation}

Substituting \eqref{eq:Lin_single_power_series_H} into \eqref{eq:readout}, we obtain
\begin{equation}
\hat{y}(x,\mathbf{w})
=\sum_{\bar{k}=0}^{\infty}\gamma_{\bar{k}}(\mathbf{w})\,x^{\bar{k}},
\label{eq:Lin_single_power_series_y}
\end{equation}
where 
\begin{equation}
\gamma_{\bar{k}}(\mathbf{w})
\triangleq
\Re\!\left\{
\left(\frac{1}{N_\mathrm{R}}\mathbf{1}_{N_\mathrm{R}}\right)^{\top}
\mathbf{E}_{\bar{k}}(\mathbf{w})\,\mathbf{1}_{N_\mathrm{T}}
\right\}.
\label{eq:gamma_n_def}
\end{equation}
Consequently, in the presence of MC, linear encoding generates higher-order powers of $x$ through the repeated interactions with the PM. Stronger MC typically leads to a slower decay of the series in \eqref{eq:Lin_single_power_series_y}, allowing more higher-order terms to contribute significantly. However, as in Sec.~\ref{sec_discuss_A}, the coefficients cannot be chosen freely in our WPNNs; they are parameterized by ${\mathbf{w}}$ as seen in (\ref{eq:gamma_n_def}).

\subsubsection{Multi-Layer WPNN with Linear Encoding and No MC}
\label{sec_discuss_Lin_multilayer_noMC}

Based on (\ref{eq_linear_encoding}) and under the assumption of $\mathbf{S}_\mathcal{SS}=\mathbf{0}$ (which implies the absence of MC), (\ref{eq1}) collapses to
\begin{equation}
\mathbf{H}^{(\ell)}(x,\mathbf{w}^{(\ell)})
=
\mathbf{S}_{\mathcal{RT}}
+x\,\mathbf{\Theta}^{(\ell)}(\mathbf{w}^{(\ell)}),
\label{eq:Lin_layer_affine_noMC}
\end{equation}
where $\mathbf{\Theta}^{(\ell)}(\mathbf{w}^{(\ell)})
\triangleq
\mathbf{S}_{\mathcal{RS}}\,\mathbf{\Omega}^{(\ell)}(\mathbf{w}^{(\ell)})\,\mathbf{S}_{\mathcal{ST}}$.
It follows that
\begin{equation}
\mathbf{b}^{(L)}(x,\mathbf{W})
=
\left[
\prod_{m=1}^{L}\left(\mathbf{S}_{\mathcal{RT}}+x\,\mathbf{\Theta}^{(m)}(\mathbf{w}^{(m)})\right)
\right]\mathbf{1}_{N_\mathrm{T}}.
\label{eq:Lin_multilayer_prod}
\end{equation}
Since each factor in \eqref{eq:Lin_multilayer_prod} is affine in $x$, the product expands into a finite polynomial in $x$ of degree at most $L$:
\begin{equation}
\hat{y}(x,\mathbf{W})
=
\delta_{0}(\mathbf{W})+\sum_{k=1}^{L}\delta_{k}(\mathbf{W})\,x^{k},
\label{eq:Lin_multilayer_poly_y}
\end{equation}
where the real-valued coefficients $\delta_k(\mathbf{W})$ are induced by the readout \eqref{eq:readout} applied to the corresponding coefficient matrices of the product expansion in \eqref{eq:Lin_multilayer_prod}. As before, the coefficients are parametrized by $\mathbf{W}$ but cannot be chosen freely.
Thus, even without MC, depth $L$ explicitly increases the maximal polynomial degree available to the WPNN.

\section{Conclusion}
\label{sec:conclusion}

To summarize, we have presented the first systematic study of the expressivity of PM-based WPNNs with structural input encoding. Our focus was on the respective roles of encoding non-linearity, structural non-linearity induced by MC, and network depth. We used a rigorous MNT-based model whose parameters were extracted from full-wave simulations of a compact 100-element D-band PM-parametrized cavity. Within this framework, we introduced and compared two multilayer architectures: one with shared weights across layers and one with independent weights per layer. This comparison clarified that depth can improve approximation capability even when it is not associated with additional trainable parameters. Across controlled non-linear regression tasks of increasing difficulty, our results showed that phase encoding is markedly more expressive than the considered linear encoding. In particular, with phase encoding and strong MC, a single-layer WPNN already achieved very good approximation performance. This point is practically important because shallow WPNNs are especially attractive for experimental implementations, where reducing the number of layers directly relaxes constraints on footprint and hardware complexity. Our results further showed that additional layers can partly compensate for reduced structural non-linearity. Overall, our findings provide a physics-consistent picture of how encoding choice, MC strength, and WPNN depth jointly shape the expressive power of PM-based WPNNs. They also offer concrete design guidance for future experimental implementations, encouraging careful choices of encoding function and experimental platforms with strong MC.

Looking forward, on the conceptual side, future work can move beyond scalar regression toward matrix-valued function approximation, where ``multivariate non-linearity'' will matter~\cite{savinson2025universality}. Structural non-linearity and depth naturally contribute to the mixing of different inputs, and encoding non-linearity can do so if the encoding function is not simply applied element-wise (which it usually is). 
Furthermore, future work can additionally account for a decoding function that may be non-linear, for instance, due to quantized intensity-only read-out. 
Meanwhile, on the practical side, a next step consists in replacing the idealized encoding functions considered in this paper by encoding functions compatible with realistic tunable lumped elements such as commercial varactors, whose many-bit tuning typically traces a discretized, non-trivial path in the complex plane. More generally, experimental validation is an important goal for future work. Recent works on experimentally estimating proxy parameters for the PM-parametrized-cavity model considered here~\cite{sol2024experimentally,del2025experimental,ambiguityaware,reduced_rank} will enable model-based training of experimental prototypes of the WPNN architectures considered in our work.

\bibliographystyle{IEEEtran}

\end{document}